%% file: LAr_LD_NIM_v4.tex
\pdfoutput=1
\documentclass[3p, number,sort&compress]{elsarticle}

\usepackage{graphicx,color,float}
\usepackage{amsthm}
\usepackage{amsmath}
\usepackage{lineno}
\journal{Nucl. Instru. \& Meth. A}
\usepackage{hyperref}
\usepackage[normalem]{ulem}

\begin{document}
\newcommand {\T} {{\cal T}}
\newcommand {\DC} {{\delta \chi^2}}
\newcommand {\A} {{A}}
\newcommand {\al} {c} 
\newcommand {\bl} {b}
\newcommand {\erf} {\text{erf}}
\newcommand {\ttheta} {\theta} 
\newcommand {\x}  {{\theta_{\text{true}}}}
\newcommand {\y}  {{\theta_{\min}}}
\newcommand {\Tm} {{\Theta_{\min}}}
\newcommand {\Tt} {{\Theta_{\text{true}}}}
\newcommand {\Eta} {{\cal H}}

\newcommand {\hmu} {\hat{\mu}}
\newcommand {\hnu} {\hat{\nu}}
\newcommand {\hpi} {\hat{\pi}}
\newcommand {\htau} {\hat{\tau}}
\newcommand {\hth} {\hat{\theta}} 
\newcommand {\dg} {\text{diag}}
\newcommand {\p} {p}
\newcommand {\s} {s}
\newcommand {\ba} {{\bf a}}
\newcommand {\bb} {{\bf b}}
\newcommand {\bc} {{\bf c}}
\newcommand {\0} {{\bf 0}}
\newcommand {\f} {{\bf f}}
\newcommand {\e} {{\bf e}}
\newcommand {\B} {{B^*}}
\newcommand {\itau} {M^{-1}}
\newcommand {\mtau} {{M}}

\newcommand {\rf} {\text{ref}}
\newcommand {\tr} {\text{true}}
\newcommand {\NH} {{H_0}}
\newcommand {\IH} {{H_1}}

\newcommand {\Prob} {\text{Prob}}
\newcommand {\E} {\text{E}}
\newcommand {\Var} {\text{Var}}
\newcommand{\ind}{\stackrel{\text{indep}}{\sim}}
\newcommand{\cas}{\buildrel \text{a.s.} \over \longrightarrow}
\newcommand{\casleft}{\buildrel \text{a.s.} \over \longleftarrow}
\newcommand{\cd}{\buildrel d \over \rightarrow}
\newcommand{\cp}{\buildrel P \over \longrightarrow}
\newcommand{\apsim}{\buildrel \text{approx.} \over \sim}

\newtheorem{lemma}{Lemma}

\newcommand {\Tbarn} {\overline{\Delta T_{H_0}}}
\newcommand {\Tbara} {\overline{\Delta T_{H_1}}}
\newcommand {\n} {{n}}
\newcommand {\Ni} {{N_i}}
\newcommand {\nutn} {\eta} 
\newcommand {\nuta} {\zeta} 
\newcommand {\nutnT} {\eta^0} 
\newcommand {\nutaT} {\zeta^0} 
\newcommand {\muT} {\mu^0} 
\newcommand {\nuT} {\nu^0} 
\newcommand {\piT} {{\pi^{0}}} 
\newcommand {\tauT} {\tau^0} 
\newcommand {\tmu} {\tilde{\mu}}
\newcommand {\tnu} {\tilde{\nu}}

\newcommand {\hnutn} {\hat{\nutn}}
\newcommand {\hnuta} {\hat{\nuta}}

\newcommand {\paratheta} {\beta} 
\newcommand {\PT} {B}

\newcommand {\D} {\overline{D}}

\def\Journal#1#2#3#4{{#1} {\bf #2}, #3 (#4)}
\def\NCA{\rm Nuovo Cimento}
\def\NPA{{\rm Nucl. Phys.} A}
\def\NIM{\rm Nucl. Instrum. Methods}
\def\NIMA{{\rm Nucl. Instrum. Methods} A}
\def\NPB{{\rm Nucl. Phys.} B}
\def\PLB{{\rm Phys. Lett.}  B}
\def\PRL{\rm Phys. Rev. Lett.}
\def\PRD{{\rm Phys. Rev.} D}
\def\PRC{{\rm Phys. Rev.} C}
\def\ZPC{{\rm Z. Phys.} C}
\def\JPG{{\rm J. Phys.} G}

\begin{frontmatter}
\title{Measurement of Longitudinal Electron Diffusion in Liquid Argon}

\author[physics]{Yichen Li\corref{correspond}}
\cortext[correspond]{Corresponding author}
\ead{yichen@bnl.gov}
\author[instrument]{Thomas Tsang}
\author[physics]{Craig Thorn}
\author[physics]{Xin Qian}
\author[physics]{Milind Diwan}
\author[physics]{Jyoti Joshi}
\author[physics]{Steve Kettell}
\author[physics]{William Morse}
\author[instrument]{Triveni Rao}
\author[physics]{James Stewart}
\author[physics]{Wei Tang}
\author[physics]{Brett Viren}
\address[physics]{Physics Department, Brookhaven National Laboratory, Upton, NY, USA}
\address[instrument]{Instrumentation Division, Brookhaven National Laboratory, Upton, NY, USA}

\begin{abstract} 
We report the measurement of longitudinal electron diffusion coefficients in liquid
argon for electric fields between 100 and 2000 V/cm with a gold photocathode as a bright 
electron source. The measurement principle, apparatus, and data analysis are described. 
In the region between 100 and 350 V/cm, our results show a discrepancy with the previous
measurement~\cite{Cennini}. In the region between 350 and 2000 V/cm, our results 
represent the world's best measurement. Over the entire measured electric field range, our 
results are systematically higher than the calculation of Atrazhev-Timoshkin~\cite{Atrazhev}.
The quantum efficiency of the gold photocathode, the drift velocity and longitudinal 
diffusion coefficients in gas argon are also presented. 
\end{abstract}
\begin{keyword}
Liquid Argon Time Projection Chamber\sep Electron Diffusion\sep Electron
Drift Velocity \sep Photocathode \sep Longitudinal Diffusion \sep
Electron Mobility
\end{keyword}
\end{frontmatter}


\input{intro.tex}

\input{diff.tex}
\input{apparatus_thomas.tex}

\input{measurement.tex}

\input{systematic.tex}

\input{results.tex}

\input{summary.tex}

\section{Acknowledgments}
This material is based upon work supported by the U.S. Department of Energy, 
Office of Science, Office of High Energy Physics and Early Career Research program 
under contract number DE-SC0012704.
\input{appendix.tex}

\section*{References}
\bibliographystyle{elsarticle-num}
\bibliography{LAr_LD_NIM_v4}{}
\end{document}

%% file: intro.tex
\section{Introduction}\label{sec:intro}
The development of the liquid argon (LAr) ionization chamber, pioneered by Willis and
Radeka \cite{willis74} was an important advance in high energy calorimetry, and was quickly followed
by the development of the LAr time projection chamber (TPC) \cite{rubbia77} as a fine-grained tracking calorimeter for high
energy physics experiments. LArTPCs are now the preferred technology
for many accelerator neutrino
and dark matter experiments.
At present, two LArTPCs have been constructed and operated for neutrino
physics measurements: the ICARUS~\cite{Amerio:2004ze} detector in Italy and the
ArgoNeut~\cite{Anderson:2011ce} detector in the US. Meanwhile, two other LArTPCs 
have been constructed for dark matter searches: Darkside~\cite{Alexander:2013hia} 
and WArP~\cite{Zani:2014lea} detectors. In the near term the MicroBooNE
experiment, with a 170 ton LArTPC, has begun operation in the
US~\cite{yu}. In the future, a set of
LArTPCs will be installed at Sanford Underground Research Facility (SURF) for the Deep Underground Neutrino Experiment (DUNE) to 
search for CP violation in the lepton sector and to determine the neutrino mass 
hierarchy~\cite{Adams:2013qkq}. For the near-term neutrino 
program~\cite{Adams:2015ogl}, a three-LArTPC configuration~\cite{Antonello:2015lea} 
will be implemented at Fermilab to search for a light sterile neutrino and 
to precisely measure neutrino-argon interaction cross sections. 

LArTPCs are attractive detectors for neutrino experiments. As the most abundant noble
gas ($\sim$~1.3\% by weight) in the atmosphere, argon is commercially 
available in large quantities. The low cost and relative high density 
($\sim$~1.4 g/ml at 87 K) make LAr an ideal material for the massive 
TPCs needed for neutrino-induced rare processes. Their resulting charged particles 
transverse through the LAr and produce ionization electrons and 
scintillation light.  
Electrons in the ionization track will then drift at constant velocity along the lines of an applied electric field.
The coordinate information perpendicular to the electron drift 
direction can be determined with a high-resolution, two-dimensional charge detector 
(e.g. wire planes).
The scintillation light provides a fast indication of initial activity.
By measuring the time delay to the subsequent signal from the drifting charge, it is possible to determine the distance over which the electrons drifted.
The spatial resolution of the interaction point and the detailed topology of the subsequent particle trajectories can reach the millimeter level.
The reconstructed event topology can be used for the particle identification and, along with measuring the total amount of drifted charge, it can be used to determine the energy deposited in the detector. 
The charge collected per wire per 
unit time is closely related to the energy deposition per unit 
distance (dE/dx), which can also be used for the particle identification and the 
energy reconstruction. Therefore, the excellent signal efficiency and
background rejection that result from these characteristics make LArTPCs
ideal for neutrino experiments. 

In order to fully optimize the extraction of the intrinsic physics information
from the recorded charge signal and to properly simulate the performance of
LArTPCs, knowledge of the transport properties of electrons in LAr is essential.
In particular, the diffusion of electrons drifting in the electric field from
the point of ionization to the anode read-out plane is an important quantity
contributing to the ultimate spatial resolution of the future
long-drift-distance (up to 20 meters) detectors. The diffusion of electrons in
strong electric fields is generally not isotropic. Therefore,
    longitudinal and transverse diffusion require separate measurements. For
    most substances the diffusion in the direction of the drift field
    (longitudinal diffusion) is smaller than the diffusion in the direction
    transverse to the field (transverse diffusion).  Measurements have been
    reported previously of transverse diffusion at electric fields above 1500
    V/cm~\cite{shibamura,Derenzo:1974ji} and for longitudinal diffusion between
    100 and 350 V/cm~\cite{Cennini}.  In this paper we report a complete set of
    measurements of the longitudinal diffusion coefficient for electrons
    drifting in LAr at fields from 100 and 2000 V/cm. In Sec.~\ref{sec:diff}, we
    review the basic formalism of electron diffusion in LAr. In
    Sec.~\ref{sec:app}, we describe our experimental apparatus. In
    Sec.~\ref{sec:measurement}, we describe the procedure of data taking and
    the analysis of the raw waveforms. In Sec.~\ref{sec:systematics}, we discuss the
    systematic uncertainties in our measurement. In Sec.~\ref{sec:results}, we
    report the results of electron drift velocity and diffusion in liquid argon
    as well as gas argon. A summary is presented in Sec.~\ref{sec:summary}.

%% file: diff.tex
\section{Electron Diffusion in Liquid Argon}~\label{sec:diff}
\begin{figure}[H]
\centering
\includegraphics[width=50mm]{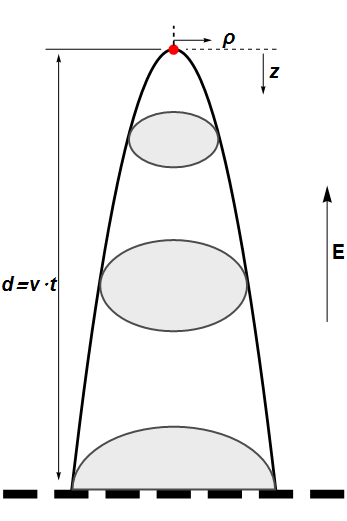}
\caption{Illustration of the diffusion process starting 
from a point source to the detection plane indicated by dashed line. The diffusion in the longitudinal direction (along the
drift direction) is generally expected to be smaller than that of the 
transverse direction. The standard deviation of the electron
swarm (indicated by the shaded ellipses in the figure at three drift
        times) is expected to 
follow Eq.~\protect\eqref{eRMS}.}
\label{fig:diffusion}
\end{figure}

When a macroscopic swarm of electrons moves through a medium 
under the influence of an electric field, three processes 
are necessary to adequately describe its time development.
They are 1) the drift velocity of the swarm centroid, 
2) the diffusional growth of the volume of the swarm, and 
3) the loss or gain of electrons in the swarm due to attachment to 
atoms or molecules in the medium or to ionization of the medium. In
liquid argon at the fields considered here, ionization does not occur.
The differential equation describing the time evolution of the electron 
density in the swarm is expressed by 
(Fick's equation~\cite{Fick1,Fick2}):
    
\begin{equation}\label{fick}
\frac{\partial n}{\partial t}=D_L\frac{\partial^2n}{\partial z^2}+D_T\left(\frac{\partial^2n}{\partial x^2}+\frac{\partial^2n}{\partial y^2}\right)-v\frac{\partial n}{\partial z}-\lambda vn,
\end{equation}
where $n\equiv n(x,y,z,t)$ is the electron charge density distribution
at position $(x,y,z)$  and time $t$.  The drift occurs with velocity $v$
in the $z$ (longitudinal) direction as shown in Fig.~\ref{fig:diffusion}. $v$ is the
drift velocity. $D_L$ and $D_T$ are the longitudinal and transverse
diffusion coefficients, respectively.  The attachment
coefficient minus the ionization coefficient is represented by $\lambda$. Since ionization does not
contribute in our case, the inverse of $\lambda$ is equal to the mean
free path.  The solution to this equation for an initial point source of
charge at the origin with a constant field in the $z$ direction is
described by a distribution function
\begin{equation}\label{ficksol}
 n(\rho,z,t)=\frac{n_0}{4\pi D_T t\sqrt{4\pi D_L t}}\exp\left(-\frac{(z-vt)^2}{4D_Lt}-\lambda vt\right)\exp\left(-\frac{\rho^2}{4D_Tt}\right),
\end{equation}
where $\rho^2=x^2+y^2$ is the transverse coordinate. This distribution
is a Gaussian function of $z$ and $\rho$ at an instant in time, but is not
a Gaussian in time at a fixed point in space. The electron
current measured on a plane perpendicular to the drift direction at a
distance $d$ from the point source (i.e. the anode) is
\begin{equation}\label{eq:e_current}
j(t)=2\pi\int\rho \cdot n(\rho, d, t) \cdot d\rho =\frac{n_0}{\sqrt{4\pi D_L t}}
    \exp\left(-\frac{(d-vt)^2}{4D_Lt}-\lambda vt\right).
\end{equation}
This function
approaches a Gaussian distribution for large $d\cdot v$
and small $D_L$, which is a reasonable approximation for our study. For this signal, the time at the peak, $t_p$, is given by:
\begin{equation}\label{eq:tpsol}
t_p=\frac{-D_L+\sqrt{D_{L}^2+d^2v(v+4D_{L}\lambda)}}{v(v+4D_{L}\lambda)},
\end{equation}
where $d$ is the drift distance. The measured
drift velocity $v_{M}$ is defined as the drift distance $d$
divided by the time at the peak $t_p$. The longitudinal diffusion 
width in time $\sigma_{L}$, defined as the standard deviation in time
of this charge distribution, can be similarly computed in terms of the true
drift velocity $v$, the diffusion coefficient $D_L$, and the
attachment coefficient $\lambda$. The resulting expression is
\begin{equation}
\sigma_L^2=\frac{2\left(d^2v_c^2-3D_L(-3D_L+\sqrt{D_L^2+d^2v_c^2})
        -dv_c(-4D_L+\sqrt{D_L^2+d^2v_c^2})\right)}{v_c^4}
\end{equation}
with $v_c=\sqrt{v(v+4D_L\lambda)}$.
For the longitudinal standard deviation we use the term diffusion time.

The expression for the measured velocity and the diffusion time 
can be expanded in a power series in $D_L$ to
        display the dependence on the parameters. The results of
        expanding both $v_M$ and $\sigma_{L}$ in powers of $D_L$ are
\begin{eqnarray}\label{eq:vdfull}
v_{M} &=&
v\left(1+\frac{D_L}{dv}+\frac{D_L^2}{2d^2v^2}+\frac{2D_L\lambda}{v}-\frac{2D_L^2\lambda^2}{v^2}+O(D_L^3,\lambda^3)\right),\nonumber\\
\sigma_L^2&=&\frac{2dD_L}{v^3}(1+\frac{17D_L}{2dv}-\frac{3D_L^2}{2d^2v^2}-\frac{6D_L\lambda}{v}-\frac{68D_L^2\lambda}{dv^2}+\frac{30D^2\lambda^2}{v^2} \nonumber 
+ O(D_L^3,\lambda^3)).
\end{eqnarray}

Provided that the dimensionless variables $\alpha=D_L/(d v)\ll$1
    and $\beta=\alpha d \lambda \ll$1, the values of $v_M$ and $\sigma_L^2$
        are well approximated by the first (zero order) terms in the
        expansions. Note that for attachment coefficients as large as
        $\lambda=1/d$, the condition that $\beta\ll 1$ is still
        satisfied. The
            full expression for $v_M$ can be inverted to give $v$ in
            terms of $v_M$, $D_L$ and $\lambda$, but the expression for
            $\sigma_L^2$ cannot be inverted. However, by substituting the
            expression for $v$ into the second series above, and
            applying Lagrange reversion~\cite{morse} to the result, we can
            obtain the series expansions for the true quantities in terms
            of the measured ones. To simplify the result, we define the
            measured diffusion coefficient as
    
   \begin{equation}
   D_{L,M}=\frac{v_M^3\sigma_{L}^2}{2d}.
   \end{equation}
 In terms of the directly measured quantities with $v_M=d/t_p$, it is
 \begin{equation}
 D_{L,M}=\frac{d^2\sigma_{L}^2}{2t^3_p}.
 \end{equation}
 With this definition, the two inverted series can be expressed in
     terms of the small dimensionless variables
     $\alpha_M=D_{L,M}/(dv_M)$ and $\beta_M=\alpha_M d\lambda$
     as:
     \begin{eqnarray}\label{eq:vinvert}
v&=&v_M(1-\alpha_M+11\alpha_M^2-\frac{427}{2}\alpha_M^3-2\beta_M+23\alpha_M\beta_M+2\beta_M^2 \nonumber \\
	&-&449\alpha_M^2\beta_M-44\alpha\beta^2+O(\alpha_M^4,\beta_M^3)),\nonumber\\
    D_L&=&D_{L,M}\left(1-\frac{23}{2}\alpha_M+\frac{449}{2}\alpha_M^2-\frac{10833}{2}\alpha_M^3+O(\alpha_M^4)\right).
 \end{eqnarray}
These expressions are used to obtain the true drift velocity and
diffusion coefficient from the measured quantities.  
   
We will show in 
Sec.~\ref{sec:systematics} that 
retaining only the first term in both these expansions results in a
negligible bias for the measurements reported here. Note that the
attenuation coefficient does not appear in the expression for the
diffusion constant: electron attachment has no effect on the measurement
of the diffusion coefficient. Even for the drift velocity, the effect of electron attachment is
still negligible: an attenuation of the electron swarm as large as
60\% ($\lambda d$=1) will only triple the already small error that
occurs when $\alpha\ll 1$ (compare the second and fifth terms in the
    series for $v$ in Eq.~\eqref{eq:vinvert}). 
The diffusion time, in terms of the true quantities, can be expressed to
the same approximation as
\begin{equation}\label{eRMS}
\sigma_{L}=\sqrt{\frac{2dD_L}{v^3}}\left(1+(\frac{17}{4}-3d\lambda)\alpha
        +\frac{1}{32}(-313-680d\lambda+336d^2\lambda^2)\alpha^2+O(\alpha^3)\right).
\end{equation}
Again, for most purposes, only the first term needs to be retained. 

In these experiments we have recorded data directly from the output of
the charge-sensitive pre-amplifier, so the measured signal is the total charge collected
on the anode of the drift cell as a function of time, which is the
integral of Eq.~\eqref{eq:e_current}.  The signal shape is then described
approximately by an
error function, and the drift time and diffusion time are determined
from the time that the signal reaches 50\% of the maximum and half of
the rise time between the 15.9\% and 84.1\% points of the signal,
respectively.  The drift velocity and the
diffusion coefficients are functions of the applied electric field.
To the extent that the higher order terms in Eq.~\eqref{eRMS} are
negligible, the longitudinal diffusion time is proportional
to the square root of the drift distance at a fixed electric field.
Fig.~\ref{fig:diffusion} illustrates the diffusion process. 

As the field approaches zero, the electrons gain so little energy from the field 
between the  elastic atomic collisions that they come to thermal equilibrium 
with the medium. In this case, the diffusion coefficients are given by 
the Einstein-Smoluchowski relation~\cite{einstein1,smoluchowski}:
\begin{equation}\label{einstein}
D=\frac{kT}{e}\mu,
\end{equation}
where $kT = 0.0075$ eV for argon at the normal boiling point (87.3 K),
$e$ is the electron charge, and $\mu$ is the electron mobility (the
        electron drift velocity $v$ per unit electric field $E$). At
low fields, when the electrons are in equilibrium with the medium, the
mobility and the electron temperature become constants.  As the field is
increased and the electrons are no longer in thermal equilibrium with
the medium, the Einstein-Smoluchowski relation is taken to define the
electron temperature.  The electrons acquire energy from the field so
that their temperature exceeds the thermal limit. The mobility decreases
and the diffusion coefficients increase. This process has been
quantified by generalized Einstein relations which obtain a quantitative
relation between diffusion and mobility for electrons in strong electric
fields using the energy dependence of the electron-atom collision cross
section. The earliest version of this class of relations for electrons,
as proposed by Wannier~\cite{wannier,robson}, has the form:
\begin{equation}\label{GER}
 \begin{aligned}
   D_T&=\frac{kT}{e}\mu,\\
   D_L&=\frac{kT}{e}(\mu+E\frac{\partial \mu}{\partial E}),
 \end{aligned}
\end{equation}
where $kT$ is the electron temperature.
The ratio of the longitudinal to the transverse diffusion coefficient is then
expressed as:
\begin{equation}\label{LTratio}
 \frac{D_L}{D_T}=1+\frac{E}{\mu}\frac{\partial \mu}{\partial E}.
\end{equation}
At low electric fields, the mobility approaches a constant, and 
we have $D_L/D_T \approx 1$. As electric field increases, if the mobility 
decreases, then $D_L < D_T$, and vice versa. 

%% file: apparatus_thomas.tex
\section{Experimental Apparatus}\label{sec:app}
Historically, there are three different techniques for the measurement
of longitudinal diffusion: un-gridded~\cite{hunter,kusano}, gridded~\cite{davies}, 
and  shuttered~\cite{takatou} drift cells. The electron source has
typically been radioactive or a photocathode driven by a ultraviolet
(UV) lamp. 
In this measurement, we chose to use a gridded drift cell because of its 
simplicity and similarity to the implementation of typical LArTPCs. 
For electron source generation, we employed an UV pulsed
laser driving a 
semitransparent metal photocathode. The short pulse width and trigger
stability of the laser allow timing information to be precisely 
synchronized. 

\begin{figure}[H]
\centering
\includegraphics[width=75mm]{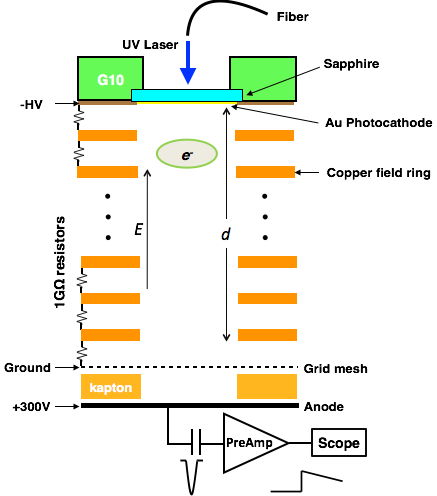}
\caption{(color online) Illustration of  setup for the electron drift velocity
and diffusion measurement. See text for further discussion. }
\label{fig:setup}
\end{figure}

Fig.~\ref{fig:setup} shows a schematic view of the experimental 
apparatus. The photocathode is composed of a 22 nm thick Au film
($\sim$~50\% UV transmission) 
thermally evaporated onto a 1 mm thick, 10 mm diameter sapphire disk. The 
sapphire disk is mounted on a G10 dielectric ring
at the top of the drift stack.
The laser source is a frequency quadrupled, 4.66 eV, 60 picosecond pulse
width Nd:YAG laser operating 
at 10 Hz repetition rate. The few $\mu$J of pulsed UV photons are  focused onto a 1-meter 
long 600 $\mu$m diameter UV fused silica fiber with an numerical aperture (NA) of 0.22.
The fiber is then coupled to a 600 $\mu$m diameter vacuum fiber feed-through followed 
by another 0.35-meter long 600 $\mu$m diameter in-vacuum UV fiber.
The latter fiber ends directly on the back surface of the sapphire substrate.
Photoelectrons are extracted into the LAr from the surface of the Au
photocathode which is back-illuminated through this 
optical fiber delivery system with a maximum full beam size of $\sim$~1 mm.
The laser energy density at the photocathode is at least an order of magnitude 
lower than the measured $\sim$~6 mJ/cm$^2$ threshold which leads to
damages of gold films. 

Negative high voltage is connected to a 200 $\mu$m thick phosphor bronze disk gasket 
that makes electrical contact to the photocathode. Ceramic screws are used to 
secure the gasket onto the G10 disk structure. Field shaping copper
electrode rings with  2 mm thickness, 21.3 mm
outside diameter, and a 15.7 mm open aperture are spaced 
5 mm apart, on center,  dictated by four G10 precision slotted rod structures. 
A series of 1 G$\Omega$ cryogenic compatible resistors make electrical contact to 
each ring and to the photocathode. A 300 $\mu$m thick Kapton ring is sandwiched 
between a +300 V biased anode disk below a grid mesh which is connected 
to the ground and to the last 1 G$\Omega$ resistor. The grid  mesh has a
wire width of 35 $\mu$m with 350 $\mu$m spacing 
and is mounted below a 250 $\mu$m thick phosphor
bronze ring. The impact of thermal contraction on the distances 
    at cryogenic temperature is negligible due to low thermal expansion 
        coefficients
        of the materials. Given this setup, various drift distances from 5 mm to 
60 mm can be arranged by adding or removing field shaping rings and
moving the cathode assembly. The field between the grid mesh and the anode is
called the anode collection field and kept at constant high value 
 independent of the field in the drift region (drift field).

A thermocouple is placed near the photocathode to monitor 
the temperature. The drift stack is housed in a 0.5 liter 
cylindrical vacuum cell capable of being pressurized up to 1.66 bar 
limited by a pressure relief valve. As shown in Fig.~\ref{fig:setup},
electron charge arriving at the anode plane is fed to a
charge sensitive pre-amplifier (IO535) developed by
Brookhaven National Laboratory which resides
outside the cryogenic cell in ambient conditions.
The pre-amplifier charge signal and its
corresponding shaped and amplified 
pulses are recorded on an oscilloscope triggered by a fast photodiode
that views a portion of the laser beam. 
The number of photoelectrons per pulse drifting through the stack to the
anode 
ranges from $10^3$ to $10^7$ at bias fields increasing from 0.01 to
$\sim$~3 kV/cm.
On the low end, this measurement is limited by amplifier noise which is
$\sim$~1350 rms electrons. With an additional 0.1 $\mu$s 
shaping amplifier, the noise charge can be further reduced to $\sim$~230 rms 
electrons. In addition, the pre-amplifier responds to noise induced
by the laser Q-switch 
particularly near the t = 0 trigger time. Therefore, in our
procedure, 64 to 512 
signal traces are averaged and subtracted from the corresponding signal trace acquired when the anode 
extraction field is off.
This data acquisition process takes about one minute per measurement at a given drift field. 

The entire system is baked to $\sim90^{\circ}\rm C$ and pumped
to $\sim10^{-8}$ Torr prior to measurements. Great attention is taken
to achieve the necessary high purity of argon gas and to avoid discharge
at the high bias fields~\cite{townsend}.  
We start with high purity commercial (99.9999 vol \%) Ar gas, which 
is then passed through a dry ice cold trap to reduce water vapor.
The Ar gas is further purified by an Oxisorb filter to a ppb impurity level, 
and then transported through stainless steel pipes and valves. The argon in the cell is not further purified by
recirculation.
Finally, Ar is liquefied by external cooling with a mixture of liquid nitrogen 
and dry ice. A CCD camera is mounted on a top viewport to observe the LAr
liquefaction process and also serves as a liquid level monitor.
To measure the work function and the quantum efficiency of the
photocathode, a fiber coupled continuous wave
UV
white light source (Energetiq LDLS) is wavelength selected by a monochromator,  fiber-coupled to the in-vacuum 
fiber feedthrough of the LAr drift cell, and the photocurrrent leaving the photocathode is measured
 by a Keithley picoammeter connected to the photocathode.

%% file: measurement.tex
\section{Measurement and Raw Waveform Analysis}\label{sec:measurement}

\begin{figure}
\centering
\includegraphics[width=90mm]{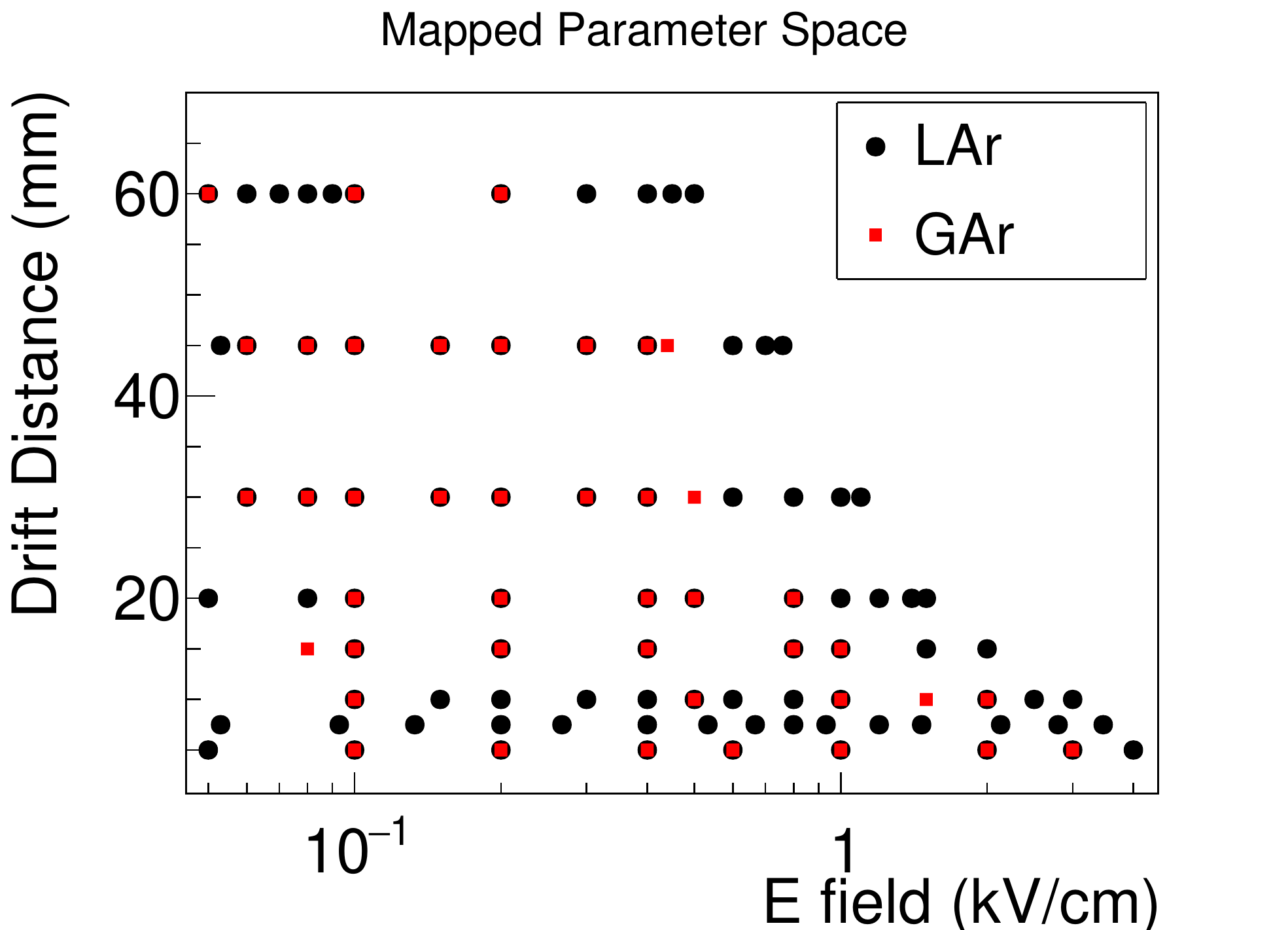}
\caption{(color online) Summary of all data taken in liquid 
argon (LAr) and gas argon (GAr) in terms 
of the drift distance and the electric field. Data above 0.1 kV/cm was used for
diffusion measurements. The boundary at high fields for each drift
distance is limited by high voltage breakdown.}
\label{fig:space}
\end{figure}

Electron drift velocity and longitudinal diffusion are determined with data
taken at various drift distances and  electric fields.
This is crucial to fully categorize systematic uncertainties. 
Changing the drift distances requires breaking the vacuum and rearranging
the drift stack. Fig.~\ref{fig:space} shows the measurement parameter space in terms of 
the drift distance vs. electric field. 
At each drift distance,
the maximum electric field that can be applied while
avoiding high voltage breakdown is $\sim$~3~kV.
It is expected that this limit may be raised with an apparatus design
that reduces sharp points causing field emission.

For each drift distance, the data taking procedure is 
\begin{enumerate}
\item measurement of electron charge versus drift field at 10$^{-8}$ Torr vacuum at room temperature, 
\item injection of ultrahigh purity Ar gas at up to 1.5 bar pressure into the test cell, 
following by charge measurement in gas Ar at room temperature, 
\item conductive cooling of the drift cell to the LAr temperature while passively maintaining the 
temperature at  87 $\pm$ 1 K and pressure of the drift cell to $1.2\pm 0.1$ bar of gas Ar,
\item charge measurements at various drift fields in LAr.
\end{enumerate}
The cooling process in step 3 generally takes $\sim$~1 hour; the measurements
in step 4 typically take several hours. It is important to note that LAr drift data 
are taken only when the thermocouple reading is 87 $\pm$ 1 K 
and the CCD camera indicates no obvious turbulence in the liquid argon.  

The pressure measurements imply a temperature  ranging from 88.2 K to
89.8 K at the surface of the LAr, in coexistence with the gas.  This is
distinctly higher than the temperature of 87 $\pm$ 1 K measured at the cathode.
We attribute this difference to a presumed temperature gradient
established in the LAr by cooling the Ar container from the bottom with the top
near room temperature.  Since no solid formation was observed during the
measurements, we assume a minimum temperature of 84 K at the bottom of
the cell.  For
these reasons we assign a mean temperature for the measurements in LAr
of 87$^{+1}_{-3}$ K. The effect of the temperature and pressure uncertainties is
discussed in Sec.~\ref{sec:drift} and Sec.~\ref{sec:diff_res}.

\begin{figure}
\centering
\includegraphics[width=90mm]{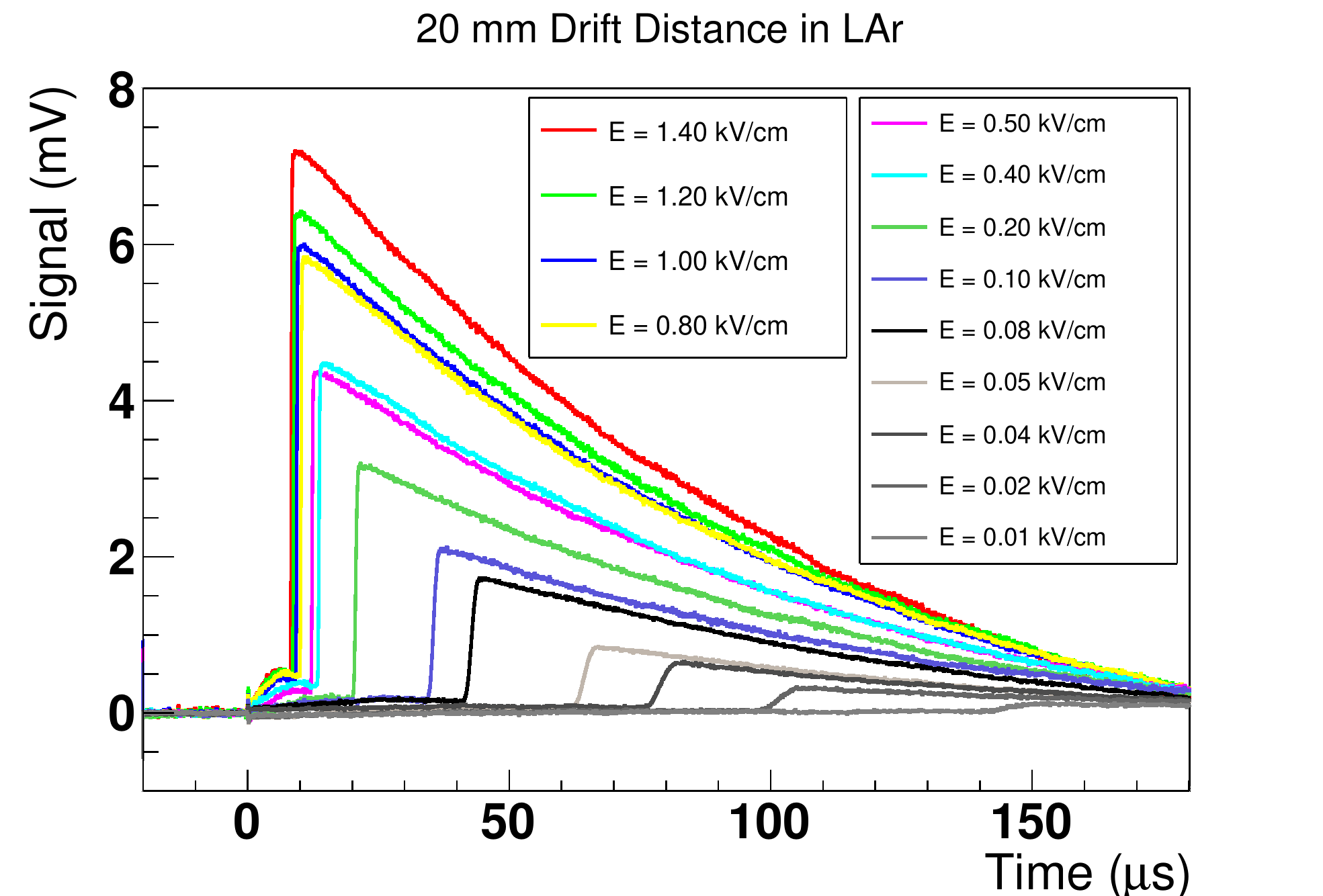}
\caption{(color online) Representative raw signals in LAr at 20 mm drift distance 
are shown for various electric fields. The 
    rising edges of the signals migrate toward the time zero with the increase of the
        electric field.
}
\label{fig:raw_signal}
\end{figure}

Fig.~\ref{fig:raw_signal} shows a representative set of pre-amplifier output data taken 
in LAr at 20 mm drift distance. Time zero is defined by the laser pulse trigger. 
The rapid rise of the signal corresponds to the time when the electron 
swarm generated at photocathode has passed
through the grid mesh. 
The position and width of the rise correspond to the drift time and diffusion time, respectively. 
The slow exponential signal decay  is
due to the 
discharge by the feedback resistor on the pre-amplifier. The slow rise prior to the onset 
of the main signal is due to some non-negligible residual charge leaving the 
anode or arriving at the cathode and does not alter the drift electron
signal shape. 
The reduction of signal height as the electric field is decreased due to the decrease in quantum efficiency at
low electric field and to the finite electron lifetime in LAr. 
The effect of the electron space charge is
negligible  because of the low charge density produced at the
photocathode (see Ref. \cite{shibamura}), which 
    will be further discussed in Sec.~\ref{sec:diff_res}.

\begin{figure}
\centering
\includegraphics[width=90mm]{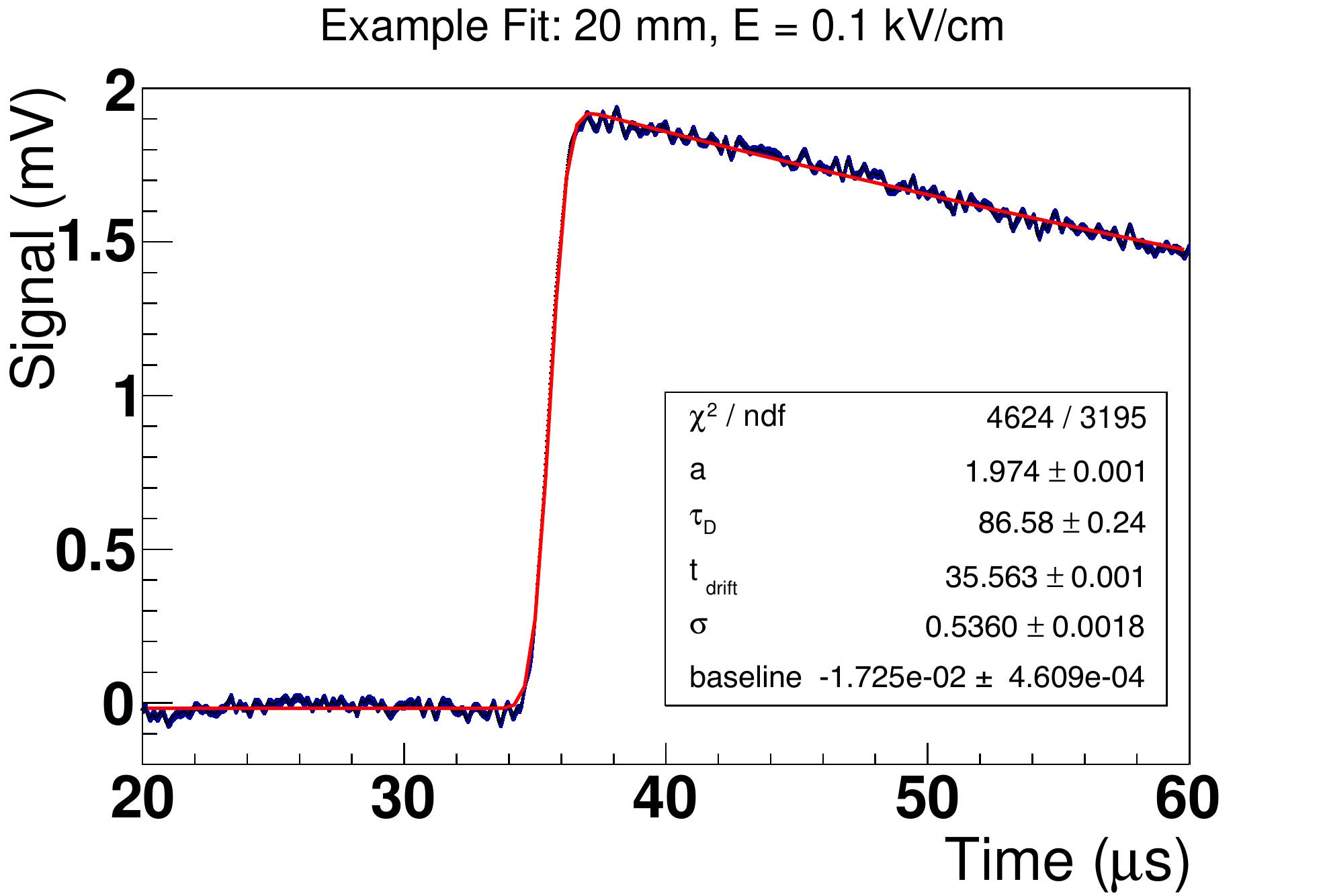}
\caption{An example fit of Eq.~\eqref{eq:fit} is shown. The uncertainty of each
data point represents the size of estimated electronic noise and is treated to 
be uncorrelated in the fit. The uncertainty of the data are enlarged 
to make the reduced chi-square $\chi^2/NDF$ to unity, which leads to 
increased uncertainties of fitted parameters.}
\label{fig:raw_fit}
\end{figure}

The raw signal is fitted with a functional form of
\begin{equation}\label{eq:fit}
f(t,\sigma,\tau_{D},a) = a \cdot e^{\frac{\sigma^2-2t\cdot \tau_{D}}{2\tau_{D}^2}} \cdot \left( 
1 + {\rm Erf}\left( \frac{-\sigma^2 + t\cdot \tau_{D}}{\sqrt{2} \tau_{D}\cdot \sigma}\right)
\right),
\end{equation}
which is a convolution of a Gaussian function approximating
 the electron swarm and a step response function with damping
representing the response of the pre-amplifier. Here, $a$ is the
normalization factor that is proportional to the number of electrons in
the swarm, $\sigma$ is the standard deviation of the Gaussian
function which describes the longitudinal diffusion, $\tau_D$
is the decay constant of damping, and $t$ is
the drift time. To accommodate the slow rise before the main signal, an additional
parameter for the baseline shift is added. A typical fit is shown in
Fig.~\ref{fig:raw_fit}. 
The uncertainty of each data point is  the size of electronic noise,
    which is estimated by the standard deviation of the baseline. This type
    of uncertainty is treated as uncorrelated during the fit, and
    statistical uncertainties of the fitted parameters are adjusted
    accordingly to make the reduced chi-square $\chi^2/NDF$
    unity. Systematic uncertainty due to the finite fitting range is estimated
    by changing the fitting range by 20\%.

%% file: systematic.tex
\section{Systematic Uncertainties}\label{sec:systematics}
The systematic errors in measuring drift velocities and diffusion
coefficients caused by neglecting the higher order terms in the expansions of
Eq.~\eqref{eq:vinvert} are negligible for our measurements. We have measured
the electron lifetime in our apparatus, and found 
that it 
ranges from 66 $\mu s$ ($\lambda$ = 0.3 cm$^{-1}$) to 530 $\mu s$
($\lambda$ = 0.009 cm$^{-1}$).
For these conditions,  the expansion variables $D_L/(dv)$
    and $D_L\lambda/v$ are at their maximum values at lowest fields and shortest drift
        distances. For 5 mm drift distance and 0.1 kV/cm field, they have values of 2.2
        $\times 10^{-4}$ and 3.4 $\times 10^{-5}$ respectively. The
        largest bias in neglecting all but the first term in the
        expansion of Eq.~\eqref{eq:vinvert} for this field and drift
        distance 
are 0.03\% for drift
velocity and 0.3\% for diffusion time.  
For fields at which the electrons are in thermal equilibrium with the
LAr ( $<\sim$ 0.1 kV/cm), the value of $\alpha$ can be computed
using the Einstein-Smoluchowski relation to be $\alpha=\epsilon_L/\Delta
V$, where $\Delta V$ is the potential drop over the drift distance and
$\epsilon_L$ is the electron energy in volts.
For higher fields and longer drift distances the errors are
    smaller.

The second systematic
uncertainty is from the impulse response function 
of the pre-amplifier.  The integration time of the pre-amplifier is
$\tau_I$. In this case, the full form of signal response function is 
\begin{equation}\label{eq:preamp}
\begin{aligned}
  f(t, \sigma, \tau_D, a, \tau_I)= a \cdot e^{\frac{\sigma^2-2 t \cdot \tau_D }{2\tau_D^2}}\left(1+{\rm Erf}\left(\frac{-\sigma^2+t \cdot \tau_D}{\sqrt{2}\tau_D\cdot \sigma}\right)\right)\\
   + a \cdot e^{\frac{\sigma^2-2t \cdot \tau_I}{2\tau_I^2}}\left(-2+ {\rm Erfc}\left(\frac{-\sigma^2+t \cdot \tau_I}{\sqrt{2}\tau_I\cdot \sigma}\right)\right),
\end{aligned}
\end{equation}
where the first term has the same functional form as that in Eq.~\eqref{eq:fit}. Due to the smallness 
of the $\tau_I$, the above equation is difficult to evaluate numerically without a computational 
demanding integration. Since the full functional form is not suitable in the large-scale fitting of 
the data, Eq.~\eqref{eq:fit} is used instead to extract the drift and diffusion time.
In this case, $\tau_I$ contributes to both the extracted drift and diffusion time (i.e. the rise time) 
of the electron signal by adding additional time, which needs to be corrected. This effect was directly calibrated by fitting 
the electron signal measured in vacuum. The contribution to the drift time and diffusion time due 
to $\tau_I$ is determined to be  31 $\pm$ 2 ns and 33 $\pm$ 5 ns, respectively. The systematic uncertainty 
due to different fitting functions (Eq.~\eqref{eq:fit} vs. Eq.~\eqref{eq:preamp}) is negligible.

\begin{figure}
\centering
\includegraphics[width=80mm]{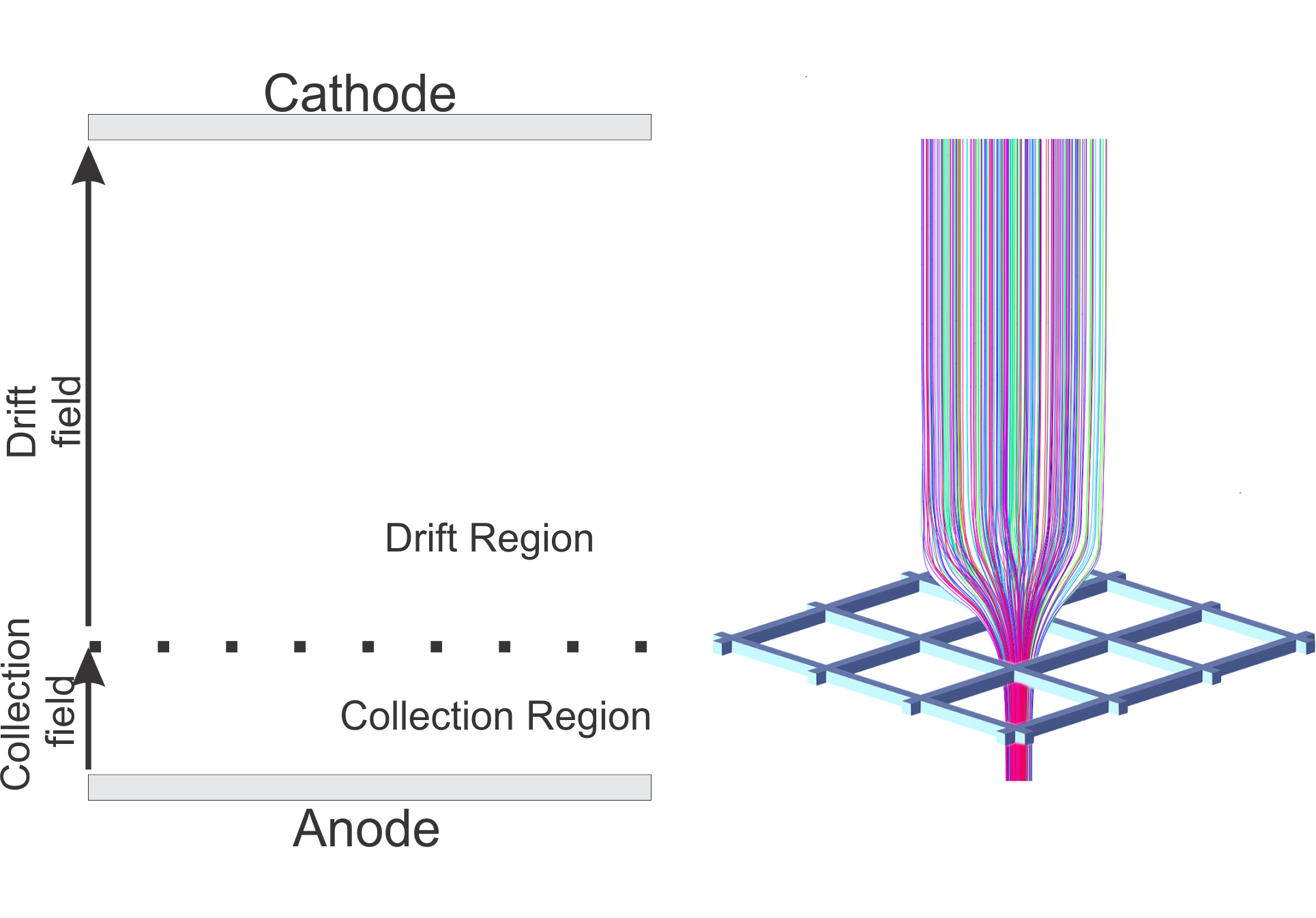}
\caption{(Left) The scheme of the electron signal simulation. The electric field is 
computed with the MAXWELL 3D in the entire geometry (i.e. including both the 
drift and collection regions) with proper boundary conditions.  
(Right) The simulated electron trajectories are plotted with different
colors respect to the 
grid mesh. The width of a grid cell is 350 $\mu$m.
The curved trajectories lead to a bias on the extraction of 
both drift time and diffusion time which are corrected in the analysis.
See text for further discussions. }
\label{fig:sim_scheme}
\end{figure}

\begin{figure}
\centering
\includegraphics[width=80mm]{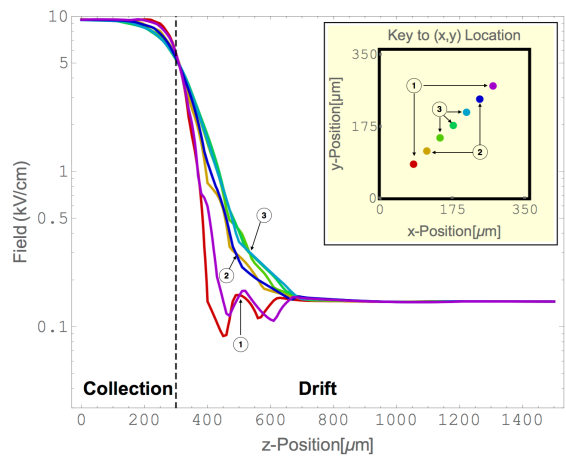}
\caption{(color online) MAXWELL 3D computed electric field close to the
    grid mesh are shown 
for a set of points along the diagonal axis of a grid cell. For the expected electric field in the 
collection and  drift region of 10 and 0.1 kV/cm, the
    achieved fields are respectively lower and higher than the expected fields due to the field leakage
    through the grid. These curves are labeled with numeric indexes by 
group on the diagonal line. }
\label{fig:efield}
\end{figure}

The largest systematic uncertainty is caused by the electron drifting from
the grid mesh to the anode. The grid mesh is installed at 
300 $\mu$m from the anode. The purpose of the grid mesh is to screen the
anode from the slow signal induced by the electrons in the drift region.
The voltage at the anode is maintained at 300 V and the grid is at
ground, creating a 10 kV/cm electric 
field in the collection region, which is much higher than applied electric field in 
the drift region (0.01 - 3 kV/cm). A perfect grid would 
separate the drift and collection regions with a constant field on both sides. 
In reality, electric field ``leaks'' through the grid from the high-field
region (i.e. the collection region) to the low-field region (i.e. the drift 
region). This effect leads to a bias in estimating the drift field. In order to evaluate this effect, the electric field in the 
drift cell is computed by MAXWELL 3D~\cite{maxwell} using a geometric model 
of the apparatus. The scheme and resulting electron trajectories
of the simulation are shown in Fig.~\ref{fig:sim_scheme}. Fig.~\ref{fig:efield} shows 
the computed electric field for a set of points along the diagonal axis of the cell. 
The electric field is largely non-uniform within -200 to +400 $\mu$m distance from 
the grid mesh.  

Another possible systematic uncertainty is the imperfect uniformity of the
electric field in our drift stack. We used MAXWELL 3D to compute the
electric fields for the full geometry including the drift stack,
         cryostat walls, and cables etc.  
Ray tracing calculations in these fields show a very small focusing of
the electron swarm as it drifts to the grid.  This causes the transverse
width of the beam distribution to decrease by about 0.08 mm, which
is significant compared to the estimated transverse diffusion especially
for short drift distances.  However, the variation in the drift time
caused by the variation in the path lengths is much smaller
fractionally, systematically increasing the longitudinal diffusion time
by a maximum of  0.31 ns compared to the measured diffusion time of 42
ns for the highest fields reported. This is a maximum systematic
increase of 0.7\%
in the longitudinal diffusion coefficient.  This systematic, one sided
error in the diffusion is typical of the calculations for all the drift
lengths.  Any other instrumental effects producing time invariant fields
also can only increase the apparent measurement of diffusion.  Among
such effects that have not been modeled are: focusing through the grid
mesh; non-co-planarity of the grid, anode, cathode and field rings;
imperfect spacing and shape of the gradient rings; charging of support
insulators; and imperfect matching of the gradient resistor values.
Including the effects of these imperfections on the central field in the
drift cell increases the assigned error for field imperfections to
$^{+2}_{-0}$\%.  This one-sided systematic error is still small compared to the
statistical error in the analysis.
 
The electron signals are simulated with a ray tracing technique that
begins with randomly 
distributed electrons within the dimension of a single grid cell (350 $\mu$m wide on each dimension). 
As shown in the right panel of Fig.~\ref{fig:sim_scheme}, 200 electrons with zero initial velocity are 
generated 5 mm above the grid plane. The electron trajectories are
solved using the computed electric field and expected electron mobility from a global fit of 
the existing world data. The electron trajectories are focused into the
center of the grid cell when electrons are close to the grid mesh,  
as shown in Fig.~\ref{fig:sim_scheme}. Therefore, 
three sources of systematic uncertainties arise due to the imperfect grid plane:
\begin{itemize}
\item The drift distance is measured to the grid mesh. The electrons
take some additional time to drift far enough toward the anode to cause
the signal rise of 50\% (the stop for the drift time measurement clock). In
addition, the trajectories starting far  from the center of the cell are
longer than the central trajectory due to the focusing as shown in the
right panel of Fig.~\ref{fig:sim_scheme}. This variation also increases the drift
time. 
This is negligible for all but the shortest drift distances and highest
drift fields as will be discussed in Section \ref{sec:drift}.
The tightly focused bundle of trajectories for low drift fields lead to a spreading of the
drift times over the grid cell, which increases the signal rise time
over that due to longitudinal diffusion only. This effect is expected to be largely
independent of the drift field. The ray tracing simulation suggests that
the magnitude of longitudinal diffusion is comparable to the electron drift time from
the grid plane to the anode plane.  Therefore, we use the expected drift
time from grid to the anode plane to estimate the contribution of
non-uniform electron trajectory to the diffusion time and assign a
conservative 100\% uncertainty in the analysis. This effect is
negligible for GAr
, but important for the LAr measurement. 
\item 
When the drift field is small, the leakage field 
leads to a significant bias on the 
magnitude of the field in the drift region. As shown in
Sec.~\ref{sec:measurement}, we limit our diffusion 
measurement to fields above 0.1 kV/cm. We further estimate the uncertainty 
due to this effect with an empirical formula (see Section
        \ref{sec:diff_res}).

\end{itemize}    

Additional systematic uncertainties include the measurement of drift distance (measured to 0.1 mm) 
and the control of experimental conditions (pressure and temperature). For LAr 
the temperature is accurate to 1 K. For GAr, it is controlled to the room
temperature, and 
the pressure is controlled to 2 mbar. The resulting uncertainties will be discussed in Sec.~\ref{sec:drift} 
and Sec.~\ref{sec:diff_res} for electron mobility and diffusion measurements, respectively.

%% file: results.tex
\section{Results}\label{sec:results}

\subsection{Quantum Efficiency of Au Photocathode}\label{sec:qe}

\begin{figure}
\centering
\includegraphics[width=90mm]{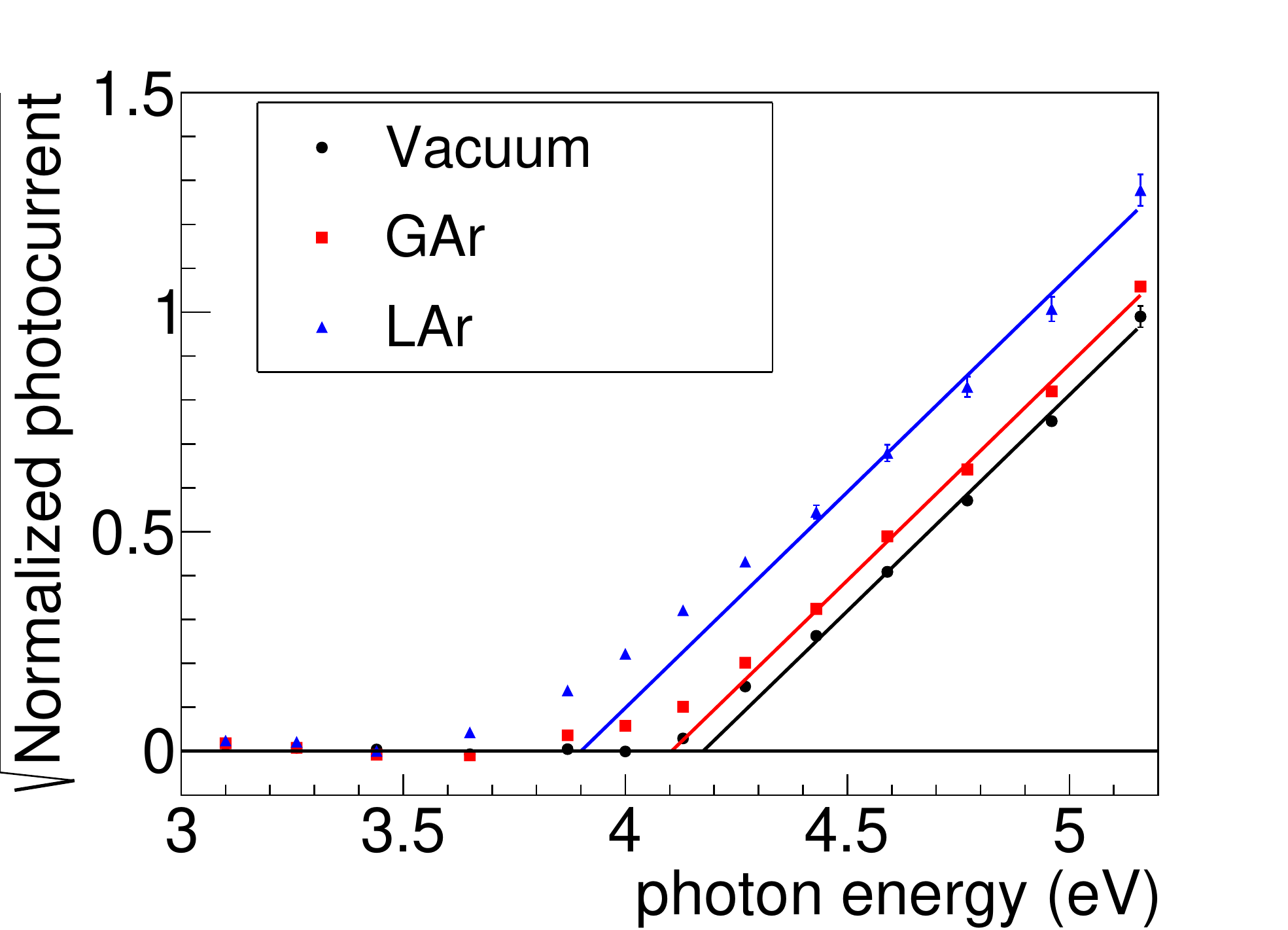}
\caption{(color online)  The square-root of photocurrents normalized to the value 
at 5.16 eV from photocathode are shown with respect to the photon energy. 
Straight line fits on the data between 4.43 and 5.14 eV are performed to 
extract the work function at y=0 intercept. Representative error bars
are shown in the fitted region.
}
\label{fig:workfunction}
\end{figure}

Several batches of Au films are measured and consistently have a work function 
of 4.17 eV, 4.10 eV, and 3.89 eV in vacuum, in  GAr, and in LAr, respectively, 
see Fig.~\ref{fig:workfunction}. The 0.28 eV lowering of work function
in LAr indicates
a negative electron affinity, which is similar to previous
measurements with Zn \cite{tauchert} and Ba \cite{halpern} photocathodes. Although the drift photoelectrons 
bear this excess kinetic energy on ejection from the cathode, they come to thermal
equilibrium through collisions
with argon atoms in a time less than 1 ns, and a distance less than a
few microns \cite{sowada}. Therefore the negative electron affinity has
no impact on drift and diffusion measurement.

The quantum efficiency (QE) is defined as the number
of electrons leaving the photocathode per number of UV photons 
irradiating the back surface of the photocathode. QEs of 
$\sim 10^{-5}$, $\sim 2\times 10^{-6}$, and $\sim 10^{-7}$ are 
obtained when the photocathode is operated in vacuum at room 
temperature, 1500 $\pm$ 2 mbar of high purity Ar gas at room temperature, 
and at LAr temperature, respectively. Despite the QE drop of nearly 
2 orders of magnitude in LAr, the photocathode provides $10^5$ 
electrons with $\sim1\mu$J of UV photons which are sufficient for the 
measurements described here.

The factor of $\sim5$ drop in the QE from vacuum 
to 1500 mbar of Ar gas is due to the increase of the potential 
barrier caused by the Ar atoms adsorbed on the photocathode, which can be 
recovered by pump down of the system and proper baking. The 
factor of $\sim50$ drop in the QE from vacuum to 
LAr is mainly caused by electron back diffusion 
to the photocathode. Although the negative electron affinity of -0.28 eV reduces
the work function of the photocathode, the field of the image charge and
the low drift velocity of electrons in LAr favor strong back diffusion so
that most photoelectrons return to the photocathode. 
A random walk simulation of the diffusion process for an electron
leaving the cathode with excess energy is presented in \cite{allen}. 
The results of this simulation suggest that when an electron 
starts on the gold-LAr interface with an electron 
energy $\sim$ 0.5 eV and propagates in a field $<$ 1 kV/cm, 
the survival rate of the photoelectrons is $\sim 3$ \%. 
This result is consistent with the QE measurements,
and also qualitatively  agrees  with other measurements on metal photocathodes
operated in LAr~\cite{tauchert}. 

\begin{figure}
\centering
\includegraphics[width=120mm]{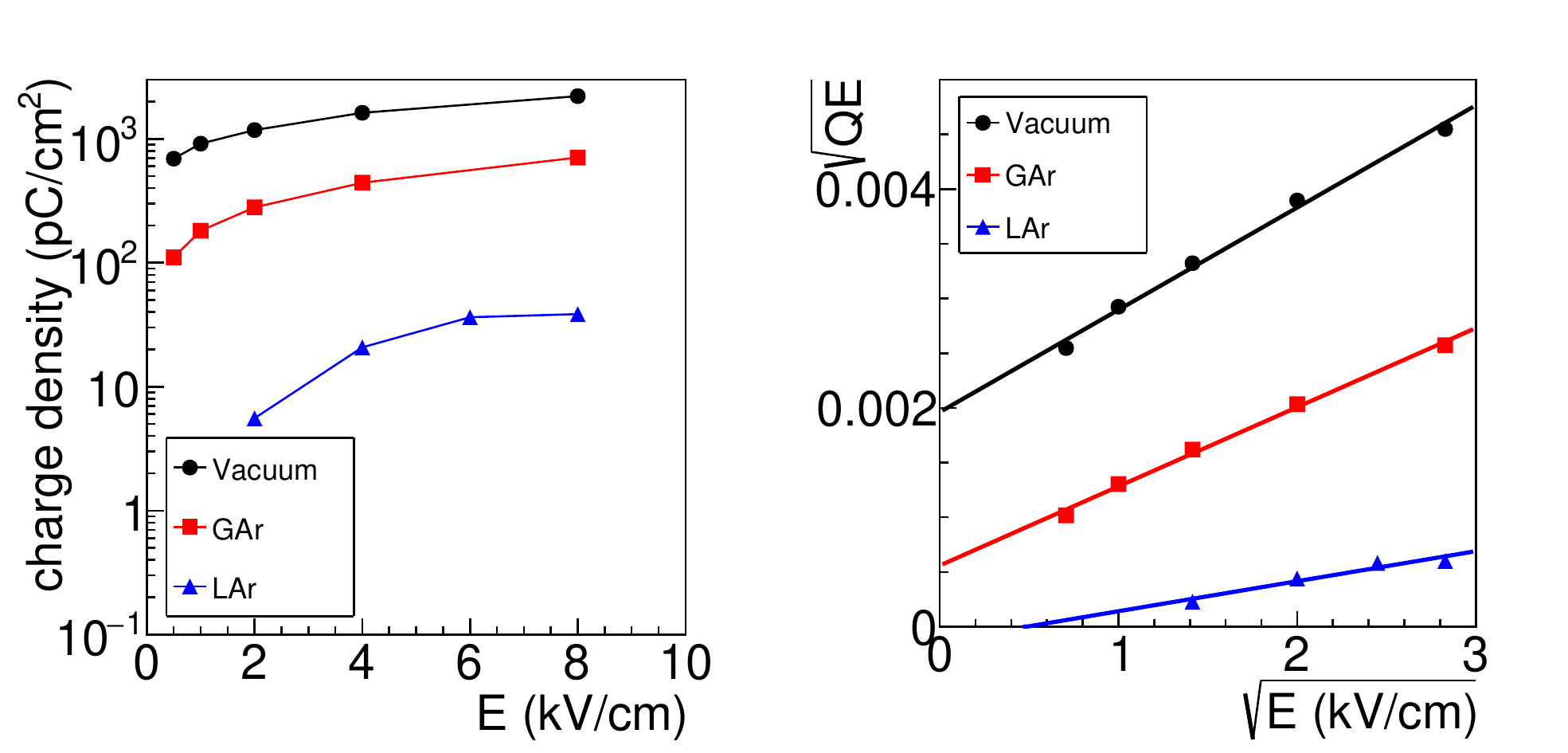}
\caption{(color online) (Left panel) Charge density from the photocathode is plotted versus the electric
field for vacuum, GAr and LAr. 
(Right panel) The square-root of the quantum efficiency is plotted against 
the square-root of the electric field. 
Linear fits of the data are shown as well. }
\label{fig:QE}
\end{figure}

Photoemission is a statistical process which can be described, 
to the first approximation, by the well-known three-step Spicer 
photoemission model~\cite{spicer2}: (i) absorption of light to excite the electrons 
from an initial state to a final state, (ii) transport of excited 
electrons to the surface, and (iii) escape of the electrons by 
overcoming a potential barrier. The QE is thus given by the Fowler law:
\begin{equation}\label{QE}
\text{QE}=k(h\nu-\phi_0+\alpha\sqrt{\beta \cdot E})^2,
\end{equation}
where $k$ is a constant that depends on the material, $\phi_0$ is the 
work function of the metal, $\alpha=\sqrt{e/4\pi\epsilon_0}$, $e$ is the 
electron charge, $\epsilon_0$ is the vacuum dielectric constant, 
$E$ is the extraction electric field, $h\nu$ is the 
photon energy, and $\beta$ is the field enhancement factor. 
The left panel of Fig.~\ref{fig:QE} plots the charge density leaving the 
photocathode versus the bias field, while the right panel of Fig.~\ref{fig:QE} 
plots $\sqrt{\text{QE}}$ against the $\sqrt{E}$. The measured QE exhibits a linear dependence 
as predicted by Fowler's law in vacuum and in low pressure of gaseous argon 
but has some deviations in liquid argon. A more quantitative evaluation
of these deviations requires measurements outside the
current scope.

\input{results_transport.tex}

%% file: results_transport.tex
\subsection{Drift Velocity in GAr and LAr} \label{sec:drift}

\begin{figure*}[htp] 
\centering \includegraphics[width=0.95\textwidth]{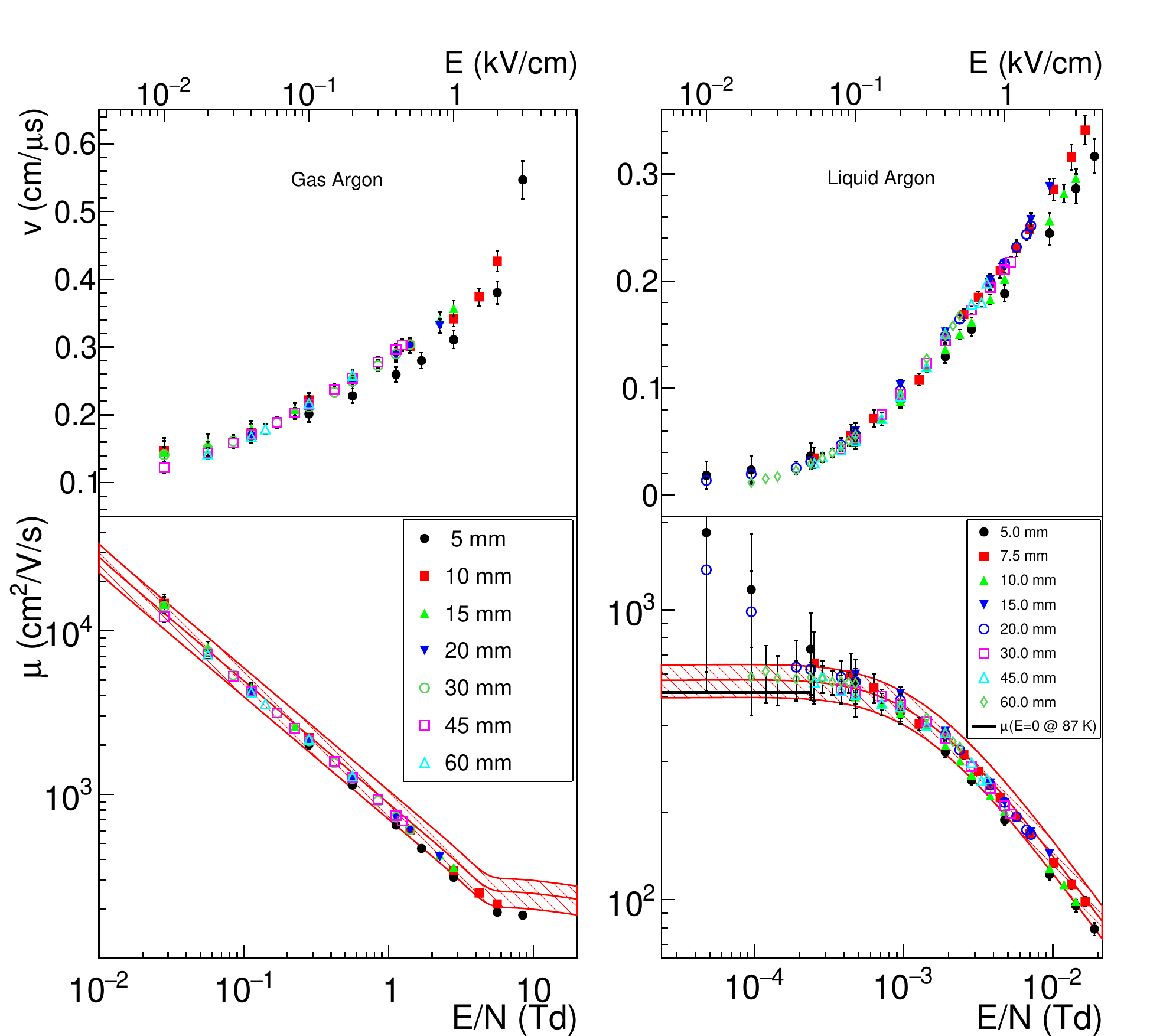}
\caption{(color online) Electron drift velocity (top panels) and electron mobility (bottom
panels) are shown with respect to the electric field for various drift distance
settings for GAr (left panels) and  LAr (right panels). 
The nominal electric field $E$ (top x-axis) and the reduced field $E/N$ (bottom x-axis) 
which is the $E$ normalized 
with the number density of argon with the unit of $Td\equiv 10^{-17}$ V$\cdot$ cm$^2$ 
are both presented for direct comparison. 
The red bands in the mobility plots (bottom panels) represent the
    global fit and their uncertainties. For LAr, the expected mobility at zero electric
    field ($\sim$~518 cm$^2$/V/s) is plotted. See text for further
    discussions.}
            \label{fig:raw_drift} 
\end{figure*}

The drift velocity and mobility are  defined as \begin{equation} v = \frac{d}{t}=\mu \cdot
E.  \end{equation} Here, $d$ is the drift distance, $t$ is the drift time, $\mu$ is the
mobility, and $E$ is the  electric field.  The drift distance $d$ is measured to an
uncertainty of 0.1 mm by the position of the anode and cathode on the drift stack.  The
drift time $t$ is calculated from the measured drift time by correcting for the time shift
due to the response of the pre-amplifier and for the time which electrons take to drift
from the grid mesh to the anode plane.  The pre-amplifier time offset is measured to be
$31 \pm 2$ ns with respect to the laser trigger from the calibration data taken in vacuum.
The drift time between the grid plane and the anode plane for GAr is calculated to be $15
\pm 5$ ns based on drift velocity data from
Refs.~\cite{Errett,pack1,brambring,wagner,wagnerEB,robertson,gar_vel,Kucukarpaci,Nakamura,pack2,Hemandez,nakamura2}.
For LAr, this correction is calculated to $65 \pm 7$ ns based on data from
Refs.~\cite{swan,Schnyders,Pruett,Miller,halpern,James,Yoshino,Gushchin,Huang,Aprile,Shinsaka,Buckley,Gonidec,Walkowiak,Amoruso}. 
The additional drift time due to grid-to-anode drift and curved
trajectories is estimated with ray tracing simulation to be $15 \pm 15$
and $65 \pm 65$ ns for GAr and LAr, respectively.  These corrections are
small compared to the measured drift time ranging from a few~$\mu$s to a few hundred~$\mu$s.
The uncertainties due to temperature and pressure variations are 
estimated to be 3\% and 5\% assuming the general~$T^{-3/2}$ temperature 
dependence of the mobility \cite{cohen}. For
LAr, this uncertainty is confirmed by repeated sets of data taken over a
period of several hours at 20 mm drift distance and 0.5 kV/cm electric
field. 

Beside these uncertainties, an additional systematic uncertainty is added due to the
leakage field as discussed in Sec.~\ref{sec:systematics}.  The leakage field leads to a
sizable difference in LAr electron mobility measured from different drift distances for
electric field below 0.1 kV/cm. Due to the difficulty in precisely predicting the size of
leakage field in all the configurations, no correction was made to our results. Instead,
conservative uncertainties for GAr and LAr were estimated by quoting a
100\% change in results when  including empirical leakage field assumed
to be uniform in the entire drift region:  
\begin{equation}\label{eq:Eleak}
        0.025 \cdot \sqrt{\frac{5 mm}{d}} 
\end{equation} 
in kV/cm. The format of the leakage field was then deduced, so that the
zero-field electron mobility in LAr then satisfied the expected results of
$\sim$518 cm$^2$/V/s. The impact of this effect is much smaller for GAr
due to much weaker electric field dependence in our region of interest
for the drift velocity.  For the diffusion measurement, we limit the
data to a region with expected electric field larger than 0.1 kV/cm,
where the leakage field effect has a smaller effect. 

Fig.~\ref{fig:raw_drift} shows electron drift velocity (top panels) and electron mobility
(bottom panels) as a function of the drift field for various drift distances for GAr (left
panels) and LAr (right panels). The nominal electric field $E$ and reduced field
$E/N$ which is the $E$ normalized with the number density of argon
with the unit of $Td\equiv 10^{-17}$ V$\cdot$ cm$^2$ are both presented
for direct comparison. This normalization 
is in particular importance for GAr comparison due to its wide range of density variation
at different pressures and temperatures.  The
 red curve and band in the bottom panels are the global fit of world
 data,  in LAr data from Ref.~\cite{swan,Schnyders,Pruett,Miller,halpern,James,Yoshino,Gushchin,Huang,Aprile,Shinsaka,Buckley,Gonidec,Walkowiak,Amoruso}, GAr data from Ref.~\cite{Errett,pack1,brambring,wagner,wagnerEB,robertson,gar_vel,Kucukarpaci,Nakamura,pack2,Hemandez,nakamura2}.

The relative uncertainties for the global fit are about 13\% and 20\%
for LAr and GAr, respectively. For GAr the extracted electron velocities
are largely consistent among different drift distances except for the 5
mm drift, which indicates additional experimental uncertainties
unaccounted for in the analysis in this setting. As shown in the
Sec.~\ref{sec:diff}, the drift time is a parameter used in the
extraction of diffusion coefficient. A 7\% additional uncertainty was
added to the measured drift time in order to take into account this
discrepancy.

For LAr, the extracted electron velocities at various drift distances are consistent with
each other. The measured velocities for 5 mm and 10 mm show a small discrepancy at high
electric field, which again indicates additional experimental uncertainties unaccounted
for in the analysis. A 5\% additional uncertainty was added to the measured drift time in
order to extract the diffusion coefficient conservatively. The comparison of the extracted LAr
electron mobility data are in excellent agreement with the global fit for electric fields
larger than 0.1 kV/cm. At electric fields smaller than 0.1 kV/cm, the leakage field effect
leads to significant systematic uncertainties. The black line represents the expected
electron mobility $\sim$~518 cm$^2$/V/s at 87 K, which is used as a guideline to estimate
the strength of leakage field. For the diffusion result discussed in the next section, we
are restricted to drift fields larger than 0.1 kV/cm.

\subsection{Longitudinal Diffusion in GAr and LAr} \label{sec:diff_res}

\begin{figure*}[htp] \centering
\includegraphics[width=0.9\textwidth]{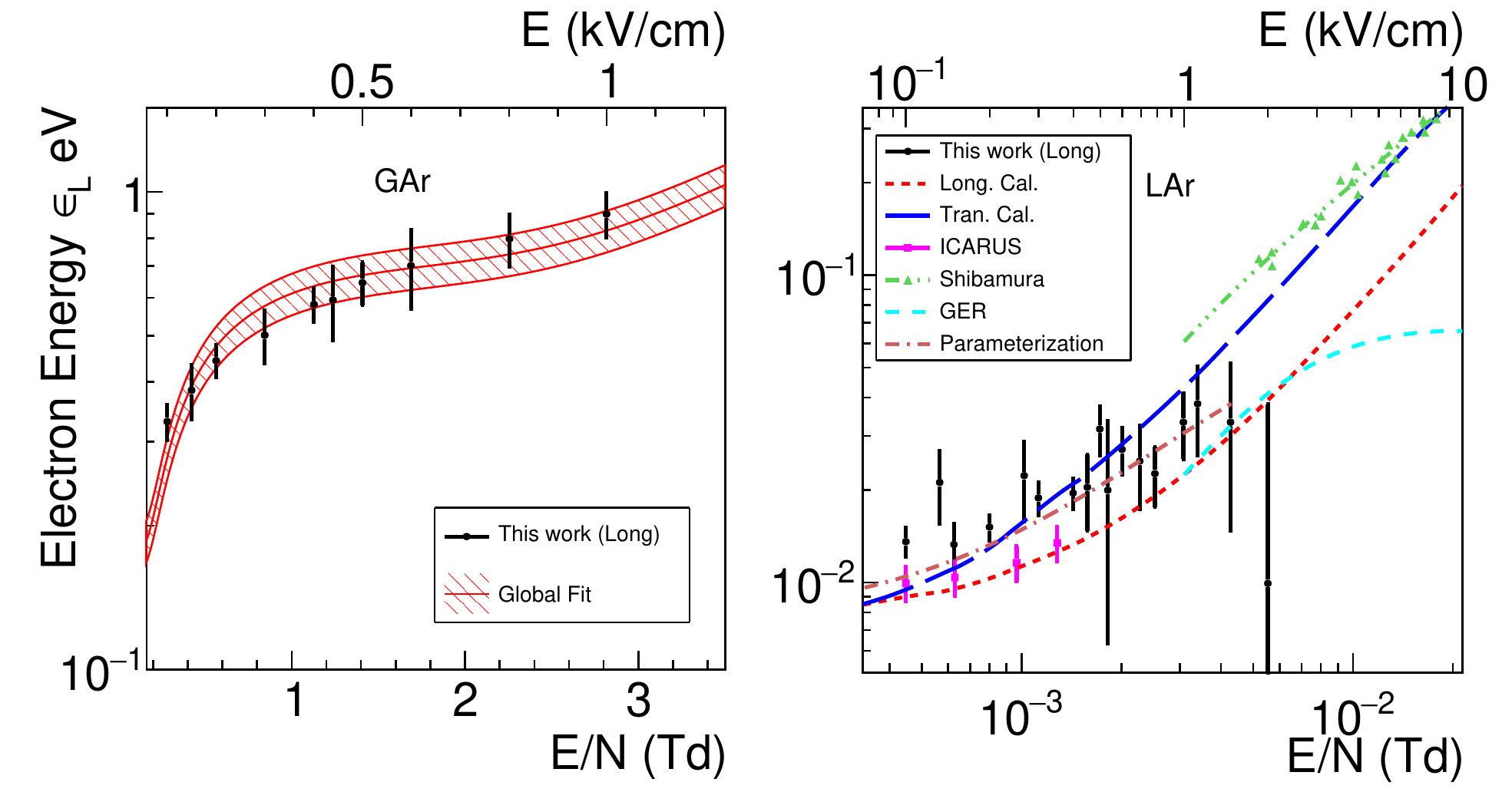}
\caption{(color online) The extracted electron energy $\epsilon_{L}$ for
    the longitudinal diffusion is plotted versus the  electric field for
        GAr and LAr. The nominal electric field $E$ (top x-axis) and reduced
        field $E/N$ (bottom x-axis) which is the $E$ normalized with the number density
            of argon with the unit of $Td\equiv 10^{-17}$
    V$\cdot$ cm$^2$ are both presented for a direct comparison.  The red band in the GAr panel
        represents the global fit and its uncertainty. For LAr, our data
        are systematically higher than the prediction of
        Atrazhev-Timoshkin~\cite{Atrazhev}. The red line represents the
        longitudinal electron energy and blue line represents the
        transverse electron energy. The magenta points are 
        deduced 
        from the measurements by ICARUS~\cite{Cennini} at low electric field
            (see text for details). The green dots 
        represent the transverse electron energy data reported by 
        Shibamura~\cite{shibamura} at high electric fields.
	The dashed cyan curve is the calculated longitudinal electron
	energy using the general Einstein relation (Eq.~\eqref{GER} and 
	Eq.~\eqref{LTratio}) from the
	interpolation (dashed green line) to the transverse electron 
        energy reported by Shibamura~\cite{shibamura}. The brown dot-dashed
        line represents the parameterization we introduced. 
        Details are presented in the appendix.} \label{fig:ar_diff} 
\end{figure*}

The longitudinal diffusion time defined in Eq. \eqref{eRMS} can be
    equivalently expressed as: 
\begin{equation}\label{eq:diff} 
\sigma_L = \sqrt{\frac{2\epsilon_{L}\cdot d}{E}} \cdot \frac{t}{d}.  
\end{equation} 
Here,  $\epsilon_{L}=eD_{L}/\mu$ is the effective energy of
electron, $D_{L}$ is the longitudinal diffusion coefficient, $\mu$
is the electron mobility, $e$ is the electron charge, $d$ is the drift
distance, $E$ is the  electric field, and $t$ is the drift time. In our
experimental setup, we expect additional contributions from systematic
errors due to the measured drift time. The contribution from the response
of the pre-amplifier is determined with the calibration data in vacuum to be
$33\pm5$ ns.  As discussed in Sec.~\ref{sec:systematics}, the
contribution due to grid-to-anode  drift and curved electron
trajectories are estimated to be $15\pm15$ and  $65\pm65$ ns for
GAr and LAr, respectively.  In this case, Eq.~\eqref{eq:diff} is
modified as: \begin{equation} \sigma^2_{L}  = \epsilon_L(E) \cdot
\frac{2 t^2}{d\cdot E} + c^2(d), \end{equation} with $c(d)$ being the
constant term that depends on the drift distance.  The longitudinal
diffusion time $\sigma_L$ and drift time $t$ depends on both the electric
field $E$ and the drift distance $d$. Therefore, the effective electron
temperature $\epsilon_L(E)$, as a function of the electric field $E$ can be
extracted through a simultaneous fit of all the results of longitudinal
diffusion and drift time by minimizing the following $\chi^2$ function:
    \begin{equation} \chi^2 = \sum_i \frac{\left(\sigma^2_{L}-
            \epsilon_L(E_{k})\cdot\frac{2 t^2}{d_{j}\cdot E_{k}}
            - c^2(d_j)\right)^2}{4\sigma_{L}^2 (\delta \sigma_{L})^2 + \frac{4t^4\cdot
        \epsilon_L^2(E_k)}{d_j^2E_k^2}\cdot \left(\frac{(\delta d_j)^2}{d_j^2} +
                \frac{(\delta E_k)^2}{E_k^2} + \frac{4(\delta t)^2}{t^2}\right)} + \sum_j
        \frac{\left(c(d_j)-c_{0}\right)^2}{(\delta c)^2}, \end{equation} 
where the notation $\delta X$ represents the uncertainty of the physical
quantity $X$.  The notation $i$ represents the $i$th data point with
electric field $E_k$ and drift distance $d_j$.  The constant term at
each drift distance is constrained around the expected value with its
uncertainty. Beside the uncertainties of measured drift time as
discussed previously, the uncertainty of the electric field is estimated
with the empirical formula in Eq.~\eqref{eq:Eleak}. Additional 
uncertainties of 1\% and 3\% are included for
GAr and LAr, respectively, to take into account the variations in
experimental conditions in terms of temperature and pressure.
This estimation was obtained assuming a linear temperature dependence 
for the electron energy.
Furthermore, due to the leakage field effect, the fit was limited to
data taken higher than 0.1 kV/cm drift field where the leakage field
effect is properly estimated as suggested by the mobility data discussed
in Sec.~\ref{sec:drift}.

The extracted electron energy $\epsilon_{L}$ for GAr and LAr are shown in
Fig.~\ref{fig:ar_diff}.  For GAr, the reduced chisquare $\chi^2/NDF = 3.3/23$ indicating
conservative systematic uncertainties.  The extracted electron temperatures are consistent
with a global fit to the world data~\cite{wagnerEB,Nakamura,pack2,Hemandez,nakamura2}. The
consistency between our data and the global fit again confirms the validity of the
analysis procedure. 

For LAr, the reduced chisquare $\chi^2/NDF = 114.6/43$ indicating additional unknown
systematic uncertainties.  In order to minimize the bias to the results
due to these uncertainties, an additional 8\% uncorrelated uncertainty
is added to each of the measured diffusion times. The new fit shown, in
the right panel of Fig.~\ref{fig:ar_diff}, gives a reduced chisquare
$\chi^2/NDF = 45.8/43$. 
The ICARUS group reports measurements for the square of
the diffusion time vs. drift time at 92 K and four fields from 100 to 350
V/cm in Fig 15. of Ref. ~\cite{Cennini}. We have fit this data to obtain
longitudinal diffusion coefficients at the four fields.  They also
report drift velocity measurements at 92 K for fields above 200 V/cm. In
order to convert the diffusion constants to effective electron
temperatures using the Einstein-Smoluchowski relation
(Eq.~\eqref{einstein}), we have obtained the corresponding mobilities
by fitting their data for mobility at 92 K (contained in Fig. 13 of
        Ref.~\cite{Cennini}) to the functional form we used for the
global fit (Eq.~\eqref{eq:globalfit} in the Appendix).  These
effective energies at 92 K are scaled to 87 K assuming a linear $T$
dependence of the effective electron energy.
This is certainly correct at low fields, where the electrons are in
thermal equilibrium with the LAr, and is probably quite accurate for the
entire range of the ICARUS data set, for which the mobility is still
close to the zero field value. Our results are higher than the converted
results from ICARUS.
\par
At high electric fields, there is no existing data on the longitudinal diffusion. 
On the other hand, transverse diffusion results exist for electric field $>$ 1.7 kV/cm 
from Shibamura~\cite{shibamura}. As discussed in 
Sec.~\ref{sec:diff}, the longitudinal electron energy can be related to the 
transverse electron energy through the general Einstein relation. Using 
Eq.~\eqref{GER} and Eq.~\eqref{LTratio} together with the electron mobility from the 
global fit (Fig.~\ref{fig:raw_drift}), the transverse electron energy results from 
Ref.~\cite{shibamura} (green points in Fig.~\ref{fig:ar_diff} and green dashed line for 
interpolation) are converted into longitudinal electron energy (dashed cyan curve in 
Fig.~\ref{fig:ar_diff}). The deduced longitudinal electron energies are consistent with 
our data at high electric fields. 

The formula of our fit to a large selection of the world's data for
electron mobility in LAr is 
presented in the Appendix. A similar parameterization is applied to
the existing data on longitudinal electron energy as shown in
Fig.~\ref{fig:ar_diff} whose formula is also presented in the Appendix.

Our results are also compared to a calculation of the transport properties of LAr by 
Atrazhev-Timoshkin~\cite{Atrazhev}, which gives the transverse and longitudinal electron 
energy as a function of field near the triple point (their Fig. 4). These calculations are shown in Fig.
\ref{fig:ar_diff}. Our data are systematically higher than this
calculation.
At zero electric field, the electron 
temperature is expected to be the same as that of the LAr. The fact that both ICARUS' and 
our data are higher than this value (0.0075 eV at 87 K) 
at low electric field 
is puzzling. In the ICARUS paper~\cite{Cennini}, an additional
correction to $D_{L}$ (and thus to the 
electron energy) caused by the space charge effect (Coulomb repulsion of the electron 
    cloud) was quoted. This correction was estimated by using an approximate model 
described in Ref.~\cite{shibamura}. The resulting electron energy from ICARUS after 
applying the correction is much lower than the LAr temperature.  We performed 
the same calculation and found that the correction for our conditions is small 
due to the lower electron density in our measurements for both drift and
collection region. The focusing of the electron trajectories through the
grid mesh causes a negligible increase in any space charge broadening
since the transit time through the grid mesh is a negligible fraction of
the total drift time and the total charge is divided among many grids. We also note that
Ref.~\cite{shibamura} declined to apply 
space charge effect they computed to their measurements. We agree with Ref.~\cite{shibamura} 
and do not include the space charge effect correction in our results. Nevertheless, we 
should comment that the space charge effect at low field with short drift distance and 
maximum space charge density could be more significant and may possibly explain the 
discrepancy between our effective energy to the theoretical calculation.
This can best investigated by measurement with different space charge
density,
    which can be achieved by changing the transverse source dimensions.

%% file: summary.tex
\section{Summary}\label{sec:summary}

LArTPCs is a key detector technology for neutrino physics in searching for new 
CP violation and determining the mass hierarchy. Understanding the performance 
of LArTPCs requires a good knowledge of electron transport in LAr. In particular,
diffusion of electrons in LArTPCs limits the spatial resolution of tracks with long 
drift distances. At 500 V/cm electric field, the longitudinal diffusion
coefficient is  $D_L=7.2$ cm$^2$/s
    from measurements, and the transverse diffusion coefficient is $D_T=$ 12.0 cm$^2$/s by extrapolating existing
        data measured at high field \cite{shibamura}. For a 3.6 meter drift, as proposed in the DUNE far detector, 
the diffusion limited resolution will be around 1.8 mm (1.1 $\mu$s in time) longitudinal 
and 2.5 mm transverse to the electric field at 500 V/cm for a track produced near the cathode. 
This implies spatial resolution for the longest drift distance will not be improved 
with signal sampling time shorter than 1 $\mu$s and wire pitch much less than 3 mm. 
For extremely long drift distances, such as the 20 meter drift distance proposed 
for the GLACIER detector, the diffusion limited resolution will be 4.2 mm 
(2.7 $\mu$s in time) longitudinal and 7.7 mm transverse to the electric field
of 500 V/cm. These values become important for distinguishing
very short tracks from point like depositions which must drift for long distances.  An important example include distinguishing supernova neutrinos from the more nearly
point-like tracks made by the $\alpha$ and $\beta$ particles from the decay of radioactive
contaminants in the LAr (such as $^{39}$Ar). Since diffusion plays 
a non-negligible role in determining signal shape in large LArTPCs, 
a good knowledge of diffusion is important in simulating the device
responses and in extracting track coordinates.

In this paper, we present our measurement of the electron mobility and longitudinal diffusion coefficients in GAr and LAr. 
Our experimental setup uses a thin gold photocathode as a bright electron source. 
The work function and quantum efficiency of the gold photocathode are
measured in vacuum, GAr, and  LAr. Our result shows that the gold photocathode is a 
good candidate as a calibration source for future large LArTPCs. 

The measured electron mobility in GAr and LAr and longitudinal electron
diffusion coefficient (expressed as 
the electron energy) in GAr, over a wide range of  electric fields,
show good agreement with fits of existing world data.
This validates our experimental approach to 
systematically measure the longitudinal electron diffusion coefficients in LAr for electric fields 
between 100 and 2000 V/cm. 
The extracted longitudinal electron energy in LAr in the region between 100 and 350 V/cm shows
a discrepancy with previous ICARUS measurement. In the region of 350 to 2000 V/cm, our results
represent the world's best measurement. Over the entire electric field range, the measured 
longitudinal electron energy is found to be systematically higher than the prediction of 
Atrazhev-Timoshkin~\cite{Atrazhev}. 

Our results are limited by the systematic uncertainties coming from i) leakage field effects due 
to imperfection of grid mesh and ii) unaccounted systematic
uncertainties in the diffusion time likely due to the control
of experimental condition and noise. A new experimental setup with
improved design of the grid mesh and better control of experimental conditions has 
been constructed. Improved measurements on the electron temperature are expected 
and will be reported. 

It is also desired to extend the measurement to transverse diffusion.
While the longitudinal diffusion expresses itself as a spread in drift time, the transverse
diffusion expresses itself as a spread in space. Therefore, the anode in the current 
apparatus is not sufficient to perform measurements of transverse diffusion. A new
anode with fine position resolution is being designed for future transverse 
diffusion measurements.

%% file: appendix.tex
\section*{Appendix}
\subsection*{Global fit on electron mobility in LAr}
The electron mobility data set used in our global data fit are from
Ref.~\cite{swan,Schnyders,Pruett,Miller,halpern,James,Yoshino,Gushchin,Huang,Aprile,Shinsaka,Buckley,Gonidec,Walkowiak,Amoruso}.
LAr temperature used in this global fit is 89 K. The data reported in
the references are all scaled to this temperature with the temperature dependence of
$T^{-3/2}$~\cite{cohen}.  The fitting function is a
rational polynomial expressed as:
\begin{equation}\label{eq:globalfit}
\mu=\frac{a_0+a_1E+a_2E^{3/2}+a_3E^{5/2}}{1+(a_1/a_0)E+a_4E^2+a_5E^3}\left(\frac{T}{T_0}\right)^{-3/2},
\end{equation}
where $E$ is the electric field  in the unit of kV/cm and $a_0 = 551.6$
cm$^2$/s is the electron
mobility at zero field with temperature of $T_0$ = 89 K, $T$ is the LAr
temperature. The fitting parameters are given in Table~\ref{tb:mupara}:
\begin{table}[H]
\centering
\caption{Function paraemeters in Eq.~\eqref{eq:globalfit}}
\label{tb:mupara}
\begin{tabular}{lllll}
\centering
$a_0$ & =& 551.6     &  &  \\
$a_1$ & =&7953.7   &  &  \\
$a_2$ & =&4440.43   &  &  \\
$a_3$ & =&4.29   &  &  \\ 
$a_4$ & =&43.63  &  &  \\   
$a_5$ & =&0.2053   &  &          
\end{tabular}     
\end{table}

\subsection*{Parameterization on effective longitudinal electron energy}
We also introduce a parameterization of the effective electron energy
for the convenience of application. Both of our data and ICARUS' data at low field
are included. The parameterization is also in a form of rational
polynomial
\begin{equation}
\label{eq:ktpara}
\epsilon_L=\frac{b_0+b_1E+b_2E^2}{1+(b_1/b_0)E+b_3E^2}\left(\frac{T}{T_1}\right)
\end{equation}
where $E$ is the electric field in the unit of kV/cm and $b_0=0.0075$ eV is the electron energy
at $T_1$ = 87 K under zero field and $T$ is the LAr temperature. The parameterization can be applied to other
temperatures with a linear temperature dependence of $T$.
\begin{table}[H]
\centering
\caption{Function parameters in Eq.~\eqref{eq:ktpara}}
\label{tb:ktpara}
\begin{tabular}{lllll}
\centering
$b_0$ & =& 0.0075     &  &  \\
$b_1$ & =&742.9   &  &  \\
$b_2$ & =&3269.6   &  &  \\
$b_3$ & =&31678.2   &  &   
\end{tabular}     
\end{table}
The longitudinal diffusion coefficients $D_L$ in the range of 0.1 to 1.5
kV/cm can thus be expressed as
defined by the Einstein relation in Eq.~\eqref{einstein}:
\begin{equation}
D_L=\frac{\mu\epsilon_L}{e}=\left(\frac{a_0+a_1E+a_2E^{3/2}+a_3E^{5/2}}{1+(a_1/a_0)E+a_4E^2+a_5E^3}\right)\left(\frac{b_0+b_1E+b_2E^2}{1+(b_1/b_0)E+b_3E^2}\right)\left(\frac{T}{T_0}\right)^{-3/2}\left(\frac{T}{T_1}\right),
\end{equation}
with the parameters given in Tables.~\ref{tb:mupara} and \ref{tb:ktpara}.